\documentclass[12pt,preprint]{aastex}

\newcommand{\bL}{\bf L}
\newcommand{\bS}{\bf S}
\newcommand{\bV}{\bf V}
\newcommand{\Msun}{M_{\sun}}
\newcommand{\ei}{e_{\rm i}}
\newcommand{\ef}{e_{\rm f}}
\newcommand{\ai}{a_{\rm i}}
\newcommand{\af}{a_{\rm f}}

\newcommand{\ma}{m_{\rm A}}
\newcommand{\mai}{m_{\rm Ai}}
\newcommand{\mb}{m_{\rm B}}
\newcommand{\mbi}{m_{\rm Bi}}

\newcommand{\Mi}{M_{\rm i}}
\newcommand{\Mf}{M_{\rm f}}

\newcommand{\VT}{V_{\rm T}}
\newcommand{\Vk}{V_{\rm k}}

\newcommand{\Pp}{P_0}
\newcommand{\Pinit}{P_{\rm init}}
\newcommand{\Pb}{P_{\rm orb}}
\newcommand{\kms}{{\rm km\,s^{-1}}}
\newcommand{\ls}{{\rm ls}}
\newcommand{\maspyr}{\rm mas\,yr^{-1}}

\begin{document}
\title{Neutron Star Kicks in Isolated and Binary Pulsars: Observational
Constraints and Implications for Kick Mechanisms}

\author{Chen Wang\altaffilmark{1}, 
  Dong Lai\altaffilmark{2,1}, 
  J. L. Han\altaffilmark{1}}
\altaffiltext{1}{National Astronomical Observatories, Chinese Academy of
  Sciences, Jia 20 Datun Road, Chaoyang District, Beijing, 100012, China;
  wangchen@bao.ac.cn, hjl@bao.ac.cn}
\altaffiltext{2}{Department of Astronomy, Cornell University, 
Ithaca, NY 14853; dong@astro.cornell.edu}

\begin{abstract}
We study observational constraints on neutron star (NS) kicks for
isolated pulsars and for neutron stars in binary systems. We are
particularly interested in the evidence of kick-spin
alignment/misalignment and its dependence on the neutron star initial
spin period.  For several young pulsars, X-ray observations of compact
nebulae showed that pulsar proper motion is aligned with the spin
direction as defined by the symmetry axis of the nebula.  We also
critically examine the measurements of the proper motion and the
projected spin axis from a large sample of pulsars with
well-calibrated polarization data. We find that among the two dozen
pulsars for which reliable measurements are available, there is a
significant correlation between the spin axis and the proper motion.
For various NS binaries, including double NS systems, binaries with
massive main-sequence star companion and binaries with massive
white-dwarf companion, we obtain constraints on the kick magnitudes
and directions from the observed orbital characteristics of the
system. We find that the kick velocity is misaligned with the the NS
spin axis in a number of systems, and the NS spin period (when
available) in these systems is generally longer than several hundreds
milliseconds. These constraints, together with spin-kick alignment
observed in many isolated pulsars, suggest that the kick timescale is
hundreds of milliseconds to 1~s, so that spin-kick alignment or
misalignment can be obtained depending on the initial spin period of
the NS. We discuss the implication of our result for various NS kick
mechanisms.
\end{abstract}


%%%%%%%%%%%%%%%%%%%%%%%%%%%%%%%%%%%%%%%%%%%%%%%%%%%%%%%%%%%%%%%%%

\section{Introduction}

It has long been recognized that neutron stars (NSs) may have received
large kick velocities at birth.  First, the measured NS velocities,
several hundreds $\kms$, are much larger than their progenitors'
velocities (e.g., Lorimer et al.~1997; Hansen \& Phinney 1997;
Arzoumanian et al~2002; Chatterjee et al.~2005; Hobbs et al.~2005; see
\S 2 below).  Second, while large space velocities can in principle be
accounted for by binary break-up (see Iben \& Tutukov 1996), many
observed characteristics of NS binaries can be explained only if there
is a finite kick at NS birth (e.g., Dewey \& Cordes 1987; Yamaoka
et al.~1993; Kaspi et al.~1996; Fryer \& Kalogera~1997; Fryer
et al.~1998; Wex et al.~1999; Stairs et al.~2003; Dewi \& van den
Heuvel 2004; Willem et al.~2004; Thorsett et al.~2005; see \S 3).  In
addition, direct observations of many nearby supernovae (e.g., Wang
2004; Leonard \& Filippenko 2004) and supernova remnants (e.g. Hwang
et al. 2004) show that supernova explosions are not spherically
symmetric, consistent with the existence of NS kicks.

While the evidence for NS kicks is unequivocal, the physical origin
remains unclear. The proposed mechanisms include hydrodynamical
instabilities in the collapsed supernova core, asymmetric neutrino
emission induced by super strong magnetic fields and post-natal
electromagnetic boost (see Lai 2004; Janka et al.~2004 and references
therein; see \S 4).  One of the reasons that it has been difficult to
spin down the kick mechanisms is the lack of correlation between NS
velocity and the other properties of NSs. The situation has changed
with the recent X-ray observations of the compact X-ray nebulae of
several young pulsars, which indicate an approximate alignment between
the pulsar proper motion and its spin axis (see Lai et al.~2001;
Romani \& Ng 2003). For a number of NS binary systems, possible
spin-kick relationship can be probed from the observed binary property
(e.g. geodetic precession; see \S 3). It is therefore useful to see
whether a consistent picture about NS kicks can be obtained from the
two sets of observational constraints.

In this paper we seek empirical constraints on the kick mechanism.  We
are particularly interested in the possible alignment/misalignment
between the kick and the angular momentum of the NS, and how such
alignment/misalignment depends on the NS initial spin period.  In \S
2, we summarize and update relevant observational data on isolated
pulsars. In addition to several young pulsars for which the spin axis
can be measured from the pulsar wind nebula, one can also constrain
the spin axis from well-calibrated polarization data. We critically
assess the information from such polarization study and demonstrate a
correlation between spin axis and proper motion for these pulsars (see
Fig.~1).  In \S 3 we discuss constraints on NS kicks in various types
of NS binaries. In addition to several well-studied NS/NS binaries, we
also consider other types of binaries, such as pulsar/main-sequence
and pulsar/white-dwarf binaries. We show that in general, the kick
direction is misaligned with the NS spin axis. We discuss the
implications of our findings in \S 4.

%%%%%%%%%%%%%%%%%%%%%%%%%%%%%%%%%%%%%%%%%%%%%%%%%%%%%%%%%%%%%%%%%
\section{Kicks in Isolated Pulsars}

By now a number of statistic studies on pulsar velocity have been
carried out (e.g., Lyne and Lorimer 1994; Lorimer et al.~1997; Hansen
\& Phinney 1997; Cordes \& Chernoff 1998; Arzoumanian et al~2002;
Hobbs et al. 2005).  These studies give a mean birth velocity
$100-500$~km\,s$^{-1}$, with possibly a significant population having
$V\gtrsim 1000$~km\,s$^{-1}$.  Arzoumanian et al.~(2002) favor a
bimodal pulsar velocity distribution, with peaks around $100~\kms$ and
$500~\kms$.  An analysis of the velocities of 14 pulsars with parallax
by Chatterjee et al.~(2005) yields a similar result (with $\sigma_v
\simeq 100~\kms$ and $300~\kms$).  Another recent study of 73 young
pulsars by Hobbs et al.~(2005) gives a mean 3D pulsar velocity of
$400~\kms$, consistent with a single Gaussian distribution.

Despite some early claims, there is currently no statistically
significant correlation between the pulsar velocity and the period or
the dipole magnetic field strength (as inferred from $P,\,\dot P$)
(e.g., Lorimer et al.~1995).  This lack of correlation is not
surprising given the large systematic error in the analysis. From a
physics point of view, it is also not surprising: (1) The observed
spin period for young pulsars is 10~ms or longer, much too slow
compared to the breakup rotation rate of a NS (period 1~ms or less);
such a slow rotation would not play any dynamically important role in
the supernova explosion.  (2) The currently observed dipole field of
pulsars is $10^{12-14}$~G, much weaker than the $10^{15-16}$~G fields
required for magnetic field to affect the explosion dynamics or
neutrino emission in proto-NSs (see \S 4).

The lack of correlation between the velocity and the other property of
pulsars has been one of the reasons that theoretical models of kicks
are not well constrained.  However, recent X-ray observation of
compact nebulae around several young pulsars has provided evidence for
spin-kick alignment; these are summarized in \S 2.1. Another way of
constraining the pulsar spin axis is through radio polarization
profile (\S 2.2).

%%%%%%%%%%%%%%%%%%%%%%%%%%%%%%%%%%%%%%%%%%%%%%%%%%%%%%%%%%%%
\subsection{Spin-Kick Correlation from Study of Pulsar Wind Nebulae}

Recent {\it Chandra} observations of the pulsar wind nebulae (PWN)
have provided evidence for spin-kick alignment for several pulsars
(e.g., Pavlov et al.~2000; Helfand et al.~2001; Lai et al.~2001; Ng \&
Romani 2004; Romani 2004; see Table 1).  In particular, the X-ray
nebulae of the Crab and Vela pulsars have a two-sided asymmetric jet
at a position angle coinciding with the position angle of the pulsar's
proper motion (Pavlov et al.~2000; Helfand et al.~2001). The symmetric
morphology of the nebula with respect to the jet direction strongly
suggests that the jet is along the pulsar's spin axis.  Analysis of
the polarization angle of Vela's radio emission corroborates this
interpretation (Lai et al.~2001; Radhakrishnan \& Deshpande 2001). Ng
\& Romani (2004) performed a systematic image analysis of pulsar wind
tori to determine the pulsar spin axis, and found that several other
cases for spin-kick alignment.

In Table 1 we list all the pulsars with the projected rotation axis
position angle $\Psi_{\rm rot}$ as determined from PWN symmetry axis,
the proper motion position angle $\Psi_{\rm PM}$, and their difference
$\mid \Delta \Psi_{\Omega \cdot v}\mid$. In addition, we list the
polarization angle $\Psi_{\rm pol}$ where available (see \S 2.2) and
its difference from the proper motion position angle $\mid \Delta
\Psi_{{\rm pol} \cdot v}\mid$. For each pulsar, we obtain the initial
spin period $P_i$ using the standard equation
\begin{equation}
\Pinit=P_0\left(1-\frac{n-1}{2}\frac{\tau}{\tau_c}\right)^{1/(n-1)},
\label{eqpi}\\
\end{equation}
where $P_0$ is the observed period, $\tau_{\rm c}$ is the
characteristic age, $n$ is the breaking index, and $\tau$ is the true
age of the system.  We summarize the key results in the following (see
Table 1).

\begin{table*}[tbh]
\begin{footnotesize}
\begin{center}
\caption{Spin and proper motion directions for young pulsars with 
the spin axis determined from PWN \label{tab1}}
\begin{tabular}{lccccccccccc}
\tableline
\tableline
PSR & $\Psi_{\rm rot}$ & $\Psi_{\rm PM}$ & 
$\mid \Delta \Psi_{\Omega \cdot v}\mid $ & 
$\Psi_{\rm pol}$ & $\mid \Delta \Psi_{{\rm pol} \cdot v}\mid$ 
& $\tau$ & $\tau_c$ & $P_0$ & $\Pinit$ \\
& deg & deg & deg & deg & deg & kyr & kyr & ms & ms \\
\tableline
B0531+21    & $124.0 \pm 0.1$ & $292 \pm 10$      & $12\pm10$ & $-60\pm$10  & 8$\pm$20  & 0.95      & 1.24 &  33.1 & 19  \\
J0538+2817  & $155 \pm 8$     & $328 \pm 4$  & $7\pm9$   &             &           & 30$\pm$4  & 618  & 143.2 & 139.7$\pm$0.5  \\
B0833-45    & $130.6 \pm 0.1$ & $301 \pm 2$       & $8.6\pm4$ & 35$\pm$10   & 86$\pm$12 & ...       & 11.4 &  89.3 & $\la$70         \\
B1706-44    & $163.6 \pm 0.7$ & $160 \pm 10$ & $3.6\pm11$ & 72$\pm$10   & 88$\pm$20 & 8.9       & 17   & 102.5 & 76$\pm$4        \\
B1951+32    & $85 \pm 5$ & $252 \pm 7$       & $13\pm9$  &             &           & 64$\pm$18 & 107  &  39.5 & 27$\pm$6        \\
J1124-5916  &                 &                   & $22\pm7$  &             &           & 2.5       & 2.9  & 135.3 & 65$\pm$20       \\
\tableline
\tableline
\end{tabular}
\end{center}
\end{footnotesize}
\end{table*}

{\it PSR B0531+21 (Crab pulsar)}. The breaking index of the Crab
 pulsar has been measured to be 2.51$\pm$0.01 between glitches (Lyne
 et al.~1993). Eq.~(1) thus gives $\Pinit=19$\,ms.  Caraveo \& Mignani
 (1999) report a {\it HST}-derived proper motion for the Crab pulsar,
 ($\mu_{\alpha}$, $\mu_{\delta}$)=($-17\pm$3, 7$\pm$3)\,$\maspyr$, or
 $\mu$=18$\pm$3\,$\maspyr$ (implying projected space velocity
 $v_\perp= 140\,\kms$) with a position angle $\Psi_{\rm
 PM}=292\arcdeg\pm10\arcdeg$.  Ng \& Romani (2004) fitted the two Crab
 tori of the PWN found in the {\it CXO} images, and gave the spin
 direction, $\Psi=124.0\arcdeg\pm0.1\arcdeg$ and
 $\Delta\Psi_{\Omega\cdot v}=12\arcdeg\pm 10\arcdeg$.

{\it PSR J0538+2817}. This 143\,ms pulsar (Lewandowshi et al.~2004) is
associated with the supernova remnant S147.  From the measured proper
motion ($67^{+48}_{-22}$\,$\maspyr$) and
the separation of the pulsar from SNR center
($2.2\times10^6$\,mas), Kramer et al.~(2003) deduce the true age
of the pulsar and the remnant, $\tau=30\pm4$\,kyr, which is a factor
of 20 times less than the pulsar's characteristic age,
$\tau_c=618$\,kyr.  This implies an initial spin period of
$\Pinit=139$\,ms.  The PWN morphology indicates the spin axis at a
position angle of $\Psi_{\rm rot}=155\arcdeg\pm8\arcdeg$ (Ng \&
Romani, 2004), which differs from the proper motion axis
($328\arcdeg\pm4\arcdeg$) by less than one $\sigma$ (Romani \& Ng
2003).

{\it PSR B0833-45}. The Vela pulsar has $P_0=89.3$\,ms and
$\tau_c=11.4$\,kyr, and is associated with the large Vela supernova
remnant.  The real age of the pulsar is not known precisely. Using
$\tau/\tau_c\ga0.5$, and $n\ga 1.5$, we obtain an upper limit of the
initial period, $\Pinit < 70$\,ms. The Vela pulsar {\it CXO} ACIS
images have a typical double tori structure, from which Ng \& Romani
(2004) deduce the polar axis direction of $\Psi_{\rm
rot}=130.6\arcdeg\pm0.1\arcdeg$. Dodson et al.~(2003) have measured
the proper motion and parallax
($\mu_{\alpha\cos\delta}=-49.68\pm0.06\,\maspyr$,
$\mu_{\delta}=29.9\pm0.1\,\maspyr$ and a distance of
$287^{+19}_{-17}$\,pc), which give transverse space velocity
$\VT=61\pm2\,\kms$, and $\Psi_{\rm PM}=301\arcdeg\pm2\arcdeg$.  This
vector lies $8\fdg6\pm 4^o$ from the fitted torus axis.

{\it PSR B1706-44}. This $P_0=102.5\,$ms pulsar has a spindown age of
17.4\,kyr (Wang et al.~2000). It is supposed on the outer edge of a
shell-type supernova remnant G343.1-2.3, implying a likely association
and the pulsar proper motion direction $\Psi_{\rm
  PM}=160\arcdeg\pm10\arcdeg$ (e.g. McAdam et al. 1993; Dodson \&
Golap 2002; Bock \& Gvaramadze 2002). A best fit of a thermal spectrum
to the X-ray emission from the SNR gives a distance of 3.1\,kpc, and
an age of 8.9\,kyr (Dodson \& Golap~2002). From eq.~(\ref{eqpi}) we
find $\Pinit$=78\,ms for $n=1.5$ and $\Pinit$=72\,ms for $n=3$.
Romani et al. (2005) fit the tori of PWN and obtain the polar axis
direction $\Psi_{\rm rot}=163.6\arcdeg\pm0.7\arcdeg$, and
$\Delta\Psi_{\Omega\cdot v}=15\arcdeg\pm 11\arcdeg$.
 
{\it PSR B1951+32}. This rapidly spinning ($P_0=39.5\,$ms) radio,
X-ray, and $\gamma$-ray pulsar is located on the edge of the unusual
SNR CTB80 (Kulkarni et al.~1988; Hobbs et al.~2004). The four epochs
between 1989 and 2000 show a clear motion for the pulsar of
25$\pm4\,\maspyr$ at a position angle $252\arcdeg\pm7\arcdeg$,
corresponding to a transverse velocity $240\pm40\,\kms$ for a distance
to the source of 2\,kpc (Migliazzo et al.~2002).  The offset between
the pulsar and the center of its associated supernova remnant implies
an age for the pulsar 64$\pm$18\,kyr, somewhat less than its
characteristic age of 107\,kyr, from which they give
$\Pinit=27\pm6$\,ms (for $n$=1.5~--~3.0). From the pulsar's polar
jets, Ng \& Romani (2004) measure the spin axis at
$\approx265\arcdeg\pm5\arcdeg$, which is 13$\arcdeg$ ($\sim1.4\sigma$)
away from the proper-motion axis.

Finally, Romani (2004) mentioned another pulsar, PSR J1124-5916, where
the misalignment angle may be measurable.  This pulsar has a period of
135\,ms and a characteristic age of 2900\,yr (Camilo et al. 2002), and
is associated with the Oxygen-rich composite supernova remnant
G292.0+1.8 (Hughes et al.~2001; Gaensler \& Wallace~2003).  The
precise age of SNR G292.0+1.8 is unknown; for estimate we use
$\tau=2500$\,yr (e. g., Gonzales \& Safi-Harb 2003).  Eq.~(\ref{eqpi})
then gives $\Pinit=65\pm20$\,ms. Romani (2004) measured the elongation
of the central PWN and compared this with the direction to the
explosion center, and obtained the misaligned angle of
$\Delta\Psi_{\Omega\cdot v}\approx22\arcdeg\pm7\arcdeg$.

We note that for three of the pulsars discussed above (see Table 1),
the spin axis can be measured from the intrinsic polarization profile
(see \S 2.2).  This gives a consistent spin-velocity angle as the
measurement from PWN, after one takes into account of the possible
orthogonal mode emission from pulsars.

%%%%%%%%%%%%%%%%%%%%%%%%%%%%%%%%%%%%%%%%%%%%%%%%%%%%%%%%%%%%%%%%%%%%%%%%%%%%%
\subsection{Spin-Kick Correlation from Polarization Study}

The polarization angle of linearly polarized emission from pulsars is
related to the dipole magnetic field geometry of the emission region
of a NS.  At the pulse center, the line of sight and the spin axis are
in the same plane as the curved magnetic field line. This provides
another constraint of the projected spin axis on the plane of the sky.
Note that radio emission from pulsars could have linear polarization
parallel or orthogonal to magnetic field. If spin-kick aligns well,
the difference between intrinsic polarization angle (IPA) and proper
motion could be either $0\arcdeg$ (for normal mode emission) or
$90\arcdeg$ (for orthogonal mode emission).

Previous investigations of spin-velocity correlation based on
polarization data have given inconsistent results. Tademaru (1977)
found some evidence for alignment from the polarization angle and
proper motion data of 10 pulsars. Morris et al.~(1979) measured IPA at
the pulse center of 40 pulsars, and claimed that the polarization
direction is either parallel or perpendicular to the proper motion
vector.  Anderson \& Lyne (1983) did not find any relation between
proper motion directions and pulsar spin axis for 26
pulsars. Deshpande et al.~(1999) checked a sample of 29 pulsars for
which they estimated IPA and proper motions, and did not find any
significant matching between IPA and proper motion.

With the new measurements of proper motions of 233 pulsars (Hobbs et
al.~2005), it is useful to re-examine the spin-velocity correlation
from polarization data available. To this end, we select normal
pulsars (with characteristic age less than a few $\times 10^7$~yr) for
which the uncertainty of position angle of the proper motion is less
than 15 degree (see ATNF pulsar catalog, Manchester et al.~2005; Hobbs
et al.~2005).  We then search the literature for observations of their
polarization properties and rotation measures. Note that many
polarization profiles were not well-calibrated in the polarization
angle, and some observations have serious (10\% or more) instrumental
effect.  In our analysis, we use high frequency ($\ga$1.4~GHz) data;
low frequency polarization observations are excluded because of the
large-uncertainties of measured polarization angle from even a small
uncertainty of Faraday rotations.  The polarization profiles we use
are mostly from Parkes observations (Qiao et al. 1995; Manchester
1971; Manchester et al. 1980; Wu, et al. 1993; van Ommen et al. 1997;
Han et al. 2005, in preparation), for which we know that their
polarization angles were well calibrated. We take the polarization
angle at the maximum sweeping rate or about the pulse center, and
calculate IPA with the observation frequency and the pulsar rotation
measures. We discard the pulsars when the errors of the calculated IPA
are greater than 20 degree. Finally, we obtain the difference between
the direction of proper motion and IPA, and remove any object with
uncertainty of the difference greater than 25 degree. After above
procedures, 24 pulsars are left (see Table 2).

In Table 2, the second and third columns give directions of proper
motion on the sky plane, $\Psi_{\rm PM}$, and IPA, $\Psi_{\rm
pol}$. The fourth columns gives the difference angle $\mid\Psi_{\rm
PM}-\Psi_{\rm pol}\mid$. The fifth and sixth columns give current
period $P_0$ and characteristic age $\tau_{\rm c}$.

\begin{table*}[htp!]
\begin{small}
\begin{center}
\caption{Spin and proper motion directions for 24 pulsars with the
spin axis determined from polarization profiles  \label{tab2}}
\begin{tabular}{lcccccl}
\tableline \tableline 
PSR & $\Psi_{\rm PM}$ & $\Psi_{\rm pol}$ & $\mid\Psi_{\rm PM}-
\Psi_{\rm pol}\mid$ & $P_0$ & $\tau_{\rm c}$ & References\\
 & deg & deg & deg & ms & yr & \\

\tableline
B0149-16   & $173.4\pm3.0 $ & $91.5 \pm9.0 $ & $81.9\pm12.0$ & $832.7 $ & 1.02e+07 & 1, 2    \\
B0531+21\tablenotemark{\dagger}   
           & $292.0\pm10.0$ & $120.0\pm10.0$ & $8.0 \pm20.0$ & $33.1  $ & 1.24e+03 & 3       \\
B0628-28   & $294.2\pm2.6 $ & $31.4 \pm3.8 $ & $82.8\pm6.4 $ & $1244.4$ & 2.77e+06 & 3, 2    \\
B0736-40   & $312.9\pm6.8 $ & $161.6\pm5.3 $ & $28.7\pm12.1$ & $374.9 $ & 3.68e+06 & 4, 3, 2 \\
B0740-28   & $277.9\pm4.4 $ & $100.8\pm3.8 $ & $2.9 \pm8.2 $ & $166.8 $ & 1.57e+05 & 4, 2    \\
B0818-13   & $169.2\pm11.9$ & $51.6 \pm4.6 $ & $62.4\pm16.5$ & $1238.1$ & 9.32e+06 & 1, 2    \\
B0823+26   & $145.9\pm1.9 $ & $72.5 \pm5.2 $ & $73.4\pm7.1 $ & $530.7 $ & 4.92e+06 & 5       \\
B0833-45   & $301.0\pm2.0 $ & $35.0 \pm10.0$ & $86.0\pm12.0$ & $89.3  $ & 1.13e+04 & 3       \\
B0835-41   & $187.3\pm7.0 $ & $97.3 \pm4.3 $ & $90.0\pm11.3$ & $751.6 $ & 3.36e+06 & 4, 2    \\
B0919+06   & $12.0 \pm0.1 $ & $137.2\pm16.4$ & $54.8\pm16.5$ & $430.6 $ & 4.97e+05 & 2       \\
B0950+08\tablenotemark{\dagger}   
           & $355.9\pm0.2 $ & $82.2 \pm3.8 $ & $86.3\pm4.0 $ & $253.1 $ & 1.75e+07 & 4, 2    \\
B1133+16   & $348.6\pm0.1 $ & $94.7 \pm4.0 $ & $73.9\pm4.1 $ & $1187.9$ & 5.04e+06 & 4, 2    \\
B1237+25   & $295.0\pm0.1 $ & $138.1\pm4.4 $ & $23.1\pm4.5 $ & $1382.4$ & 2.28e+07 & 5       \\
B1325-43   & $3.2  \pm10.4$ & $66.9 \pm11.4$ & $63.7\pm21.8$ & $532.7 $ & 2.80e+06 & 6       \\
B1426-66   & $235.9\pm8.1 $ & $51.6 \pm5.9 $ & $4.3 \pm14.0$ & $785.4 $ & 4.49e+06 & 4, 2    \\
B1449-64   & $216.9\pm2.8 $ & $131.3\pm5.9 $ & $85.6\pm8.7 $ & $179.5 $ & 1.04e+06 & 4, 2    \\
B1451-68   & $252.7\pm0.6 $ & $146.9\pm3.6 $ & $74.2\pm4.2 $ & $263.4 $ & 4.25e+07 & 4, 2    \\
B1508+55   & $225.7\pm1.1 $ & $28.5 \pm6.3 $ & $17.2\pm7.4 $ & $739.7 $ & 2.34e+06 & 5       \\
B1600-49   & $268.1\pm6.5 $ & $153.9\pm17.7$ & $65.8\pm24.2$ & $327.4 $ & 5.09e+06 & 6       \\
B1642-03   & $353.0\pm3.2 $ & $74.7 \pm6.7 $ & $81.7\pm9.9 $ & $387.7 $ & 3.45e+06 & 2       \\
B1706-44   & $160.0\pm10.0$ & $72.0 \pm10.0$ & $88.0\pm20.0$ & $102.5 $ & 1.75e+04 & 3       \\
B1857-26   & $202.8\pm0.7 $ & $161.6\pm3.8 $ & $41.2\pm4.5 $ & $612.2 $ & 4.74e+07 & 3, 2, 4 \\
B1929+10   & $65.2 \pm0.2 $ & $51.4 \pm16.2$ & $13.8\pm16.4$ & $226.5 $ & 3.10e+06 & 5       \\
B2045-16   & $92.4 \pm2.6 $ & $178.5\pm3.8 $ & $86.1\pm6.4 $ & $1961.6$ & 2.84e+06 & 4, 2    \\
\tableline
\tableline
\tablenotetext{\dagger}{Cases where the $\Psi_{\rm pol}$ have an ambiguity due to orthogonal flips.}
\tablecomments{REFERENCES: (1) Qiao et al. (1995); (2) van Ommmen et al. (1997); 
(3) Han et al. (2005); (4) Manchester et al. (1980); (5) Manchester (1971); (6) Wu et al. (1993).}
\end{tabular}
\end{center}
\end{small}
\end{table*}

\begin{figure}
\includegraphics[angle=-90, scale=0.7]{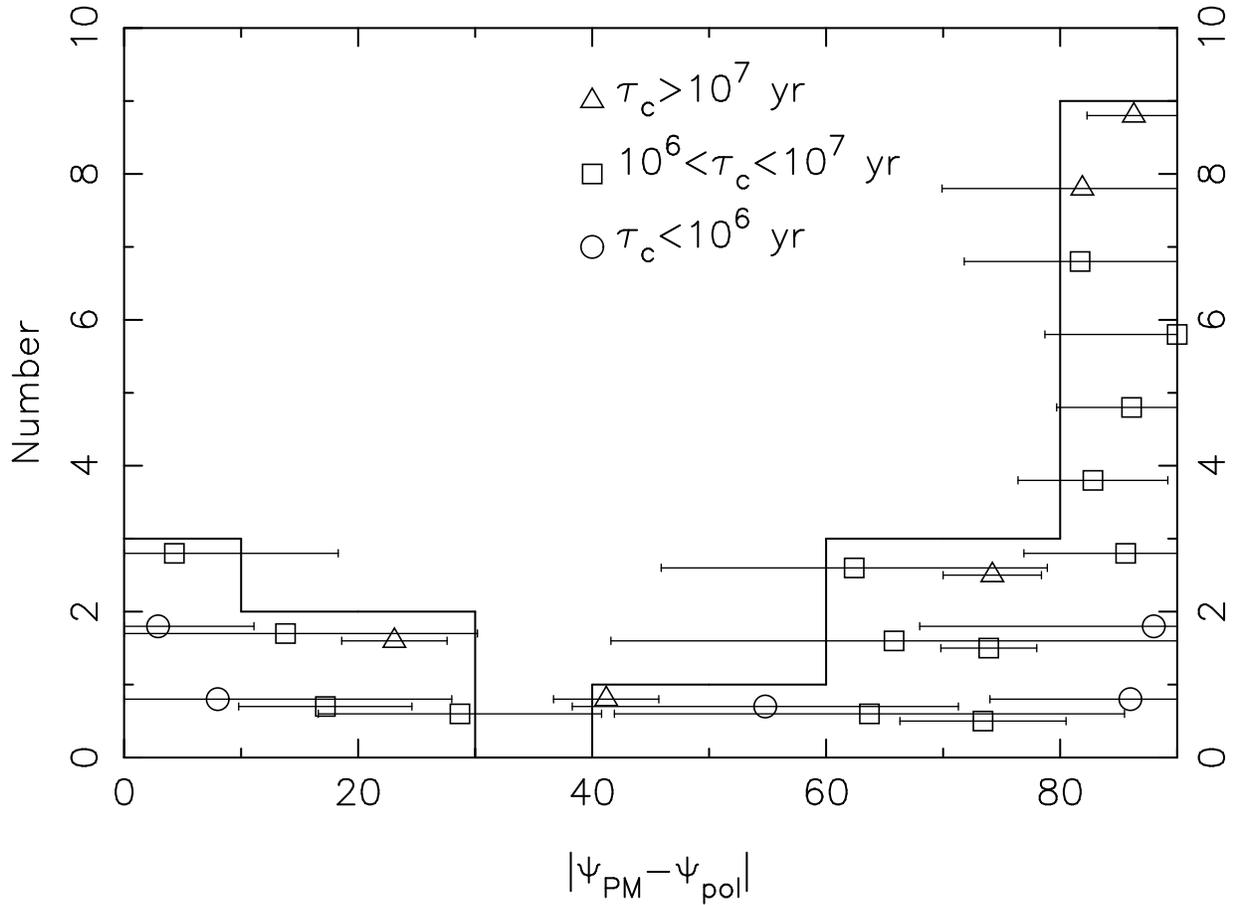}
\caption{The number distribution of the angular difference between the
proper motion and intrinsic polarization direction for 24
pulsars. Each pulsar is represented by a symbol specified by the
characteristic age of the pulsar.  }
\label{fig1}
\end{figure}

Figure 1 shows the number distribution of the angular difference $\mid
\Psi_{\rm PM}-\Psi_{\rm pol}\mid$. A significant peak appears near
$90\arcdeg$, and another peak near $0\arcdeg$ is also visible.  Since
the polarization direction is either parallel or perpendicular to the
spin axis, the data therefore indicate a significant correlation
between the spin and velocity. Given the result of PWN studies (\S
2.1), the most likely cause for the two peaks is that spin and kick
are aligned in many cases, and different pulsars prefer to emit in one
of the two orthogonal modes.  Obviously, in this interpretation,
intrinsic polarization emission favors perpendicular-mode emission for
most pulsars.

We note that for older pulsars in the sample, the proper motion may
not directly reflect the initial kick direction because of the pulsar
motion in the Galactic potential (e.g., Sun \& Han 2004). Indeed,
three of five old pulsars in Table 2 with $\tau_{\rm c}\ga10$\,Myr
(although the real age $\tau$ may be less than $\tau_{\rm c}$) show
appreciable misalignment, possibly because their proper motion
directions have been modified by the Galactic potential.

While our paper was in the final phase of preparation, we became aware
of the work by Johnston et al.~(2005b) which showed a similar
spin-velocity correlation as discussed above.

%%%%%%%%%%%%%%%%%%%%%%%%%%%%%%%%%%%%%%%%%%%%%%%%%%%%%%%%%%%%%%%%%%%%%%%%
\section{Kicks in Binary Neutron Stars}

We now study the constraint on NS kicks (both magnitude and direction)
from binary pulsar systems, including NS/NS binaries, NS/Main-Sequence
Star (MS) binaries, NS/Massive white dwarf (MWD) binaries and
high-mass X-ray binaries (HMXBs). A number of previous studies have
focused on individual systems, particularly NS/NS binaries (e.g., Wex
et al.~2000; Willems et al.~2004; Thorsett et al.~2005 and references
therein) and HMXBs (e.g. Pfahl et al.~2002). Here we consider all
systems where such kick constraints are possible.

%%%%%%%%%%%%%%%%%%%%%%%%%%%%%%%%%%%%%%%%%%%%%
\subsection{Method and Assumptions}

The basic procedure for obtaining the constraint is as follows: In the
pre-supernova (SN) binary system, we have two stars with mass 
\footnote{The subscript ``i'' specifies parameters before SN, ``f''
after SN, and ``0'' currently observed parameters.}  $\mai$ and $\mbi$
in a circular orbit (eccentricity $\ei=0$) with semi-major axis
$\ai$. Star B is a helium star ready to explode, and its mass
$\mbi=m_{\rm He}$ is constrained within the range 2.1~--~8.0\,$\Msun$
(the lower limit corresponds to the lowest mass for which a He star is
expected to form a NS instead of a white dwarf, while the upper limit
corresponds to the highest mass for which a He star is expected to
form a NS rather than a black hole; see, e.g., Fig.1 in Belczynski et
al.~2002, and Table 16.4 in Tauris \& van den Heuvel 2004).  Star A is
either a NS, a massive MS or a massive WD.  In the first case,
dynamical stability in the mass transfer from the He star to its NS
companion requires $\mbi/\mai\la 3.5$ (see Ivanova et al.~2003). As B
explodes in a SN (the explosion can be considered instantaneous
compared to the orbital period), it leaves behind a NS with mass
$\mb<\mbi$, and because of the asymmetry in the explosion or in the
neutrino emission, the NS (in its rest frame) receives a kick velocity
$\bV_k$. The angle between $\bV_k$ and the pre-SN angular momentum
$\bL_i$ is $\gamma$.  To a good approximation, star A is assumed to be
unaffected by the explosion of B (see below), i.e., its mass
$\ma=\mai$ after the SN, its post-SN velocity equals the pre-SN
velocity. Because of the mass loss and kick in the explosion, the
post-SN orbit (with semi-major axis $\af$) will in general be
eccentric ($\ef\neq 0$), and the orbital angular momentum $\bL_f$ will
be misaligned relative to $\bL_i$ by an angle $\theta$. Using angular
momentum conservation and energy conservation, we find
\begin{eqnarray}
\Vk^2 &=&\frac{G\Mf}{\af}\left[2\xi-1+\xi\eta^{-1}-2(1-\ef^2)^{1/2}\xi^{3/2}
\eta^{-1/2}\cos\theta\right],\label{eqvk}\\
\cos^2\gamma &=&\frac{\xi^2(1-\ef^2)\sin^2\theta}
    {2\xi-1+\xi\eta^{-1}-2(1-\ef^2)^{1/2}\xi^{3/2}\eta^{-1/2}\cos\theta},
\label{eqgamma}
\end{eqnarray}
where $\Mf=\ma+\mb$, $\Mi=m_{A}+\mbi$, $\eta=\Mf/\Mi$ (with $\eta<1$), 
and $\xi=\af/\ai$, which satisfies $(1+\ef)^{-1}<\xi<(1-\ef)^{-1}$.

The assumptions leading to eqs.~(\ref{eqvk})-(\ref{eqgamma}) are
fairly standard, and similar equations have been used in numerous
previous studies (e.g., Hills 1983). To obtain useful constraint on
the kick-spin correlation, we need to make further assumptions about
the rotations of the two stars:

Assumption (1): In the pre-SN binary, the assumption of circular orbit
made above is justified from the strong tidal interaction and/or mass
transfer, which also guarantee that spin angular momentum vectors of
the two stars are aligned with the orbital angular momentum vector
(i.e., $\bS_{Ai}\parallel \bS_{Bi}\parallel\bL_i$).

As an example, consider the progenitor system of a NS/NS binary: The
NS spin $\bS_{Ai}$ is expected to be aligned with $\bL_i$ because of
the mass accretion which spins up the NS. Since the He star is fully
convective, the tidal effect on the He star can be estimated using the
standard tidal friction theory. Following Zahn (1989; eq.~[21]), the
tidal circularization time is given by
\begin{eqnarray}
\frac{1}{t_{\rm circ}} = 
%-\frac{1}{e}\frac{{\rm d}e}{{\rm d}t} = 
21\frac{\lambda_{\rm circ}}{t_{\rm f}}q(1+q)\left(\frac{R_{\rm He}}{\ai}
\right)^8,
\end{eqnarray}
where $q=\ma/m_{\rm He}$ is the mass radio, $t_{\rm f}=(m_{\rm
He}R_{\rm He}^2/L_{\rm He})^{1/3}$ is the convective friction time
($m_{\rm He},\,L_{\rm He}$ are the radius and convective luminosity of
the He star), and $\lambda_{\rm circ}$ is a dimensionless average of
the turbulent viscosity weighted by the square of the tidal shear. In
the prescription of Zahn (1989), $\lambda_{\rm circ}$ can be
approximated by $\lambda_{\rm circ}\simeq 0.019\alpha^{4/3}
(1+\eta^2/320)^{-1/2}$, where $\alpha$ ($\simeq 2$) is the mixing
length parameter and $\eta=2t_{\rm f}/\Pb$ measures the timescale
mismatch. For the He star parameters, we use the result of Dewi et
al.~(2002, 2003), who studied the late stage evolution of different
types of He-star/NS binaries: (i) Case BA (mass transfer during He
core burning) with $m_{\rm He}=1.5-2.9\,\Msun$: the remnants are
heavy, degenerate CO white dwarfs (WDs). (ii) Case BB (mass transfer
during He shell burning): $m_{\rm He}=1.5-2.1\,\Msun$ produces CO or
ONe WDs, $m_{\rm He}=2.4-2.5\,\Msun$ produces ONe WDs and more massive
He stars produce NSs. (iii) Case BC (the He star fills its Roche lobe
during carbon core burning or beyond) with $m_{\rm He}=2.8-6.4\,\Msun$
producing NSs.  Table 3 lists some typical values of $m_{\rm He}$,
$R_{\rm He}$, $L_{\rm He}$ and $\Pb$ for He-star/NS binaries
considered by Dewi et al.~(2002, 2003), together with the tidal
friction time and circularization time as calculated from eq.~(4).  We
see that the tidal circularization time is about one hundred years,
and the corresponding synchronization/alignment timescale for the spin
of the He star is even shorter [by a factor $\sim (R_{\rm
He}/\ai)^2$].  These timescales are much less than the typical time of
mass transfer from the He star prior to its supernova explosion. Thus,
the assumption $\bS_{Ai}\parallel \bS_{Bi}\parallel\bL_i$ is very
reasonable\footnote{Our analyses here are based on the model of Dewi
et al. (2002, 2003), which posits that the He star fills its Roche
lobe at certain stage of its evolution. Such Roche-lobe-filling
configuration is most likely required for many systems considered here
(e.g., PSR J0737-3039 and PSR B1534+12). We note that population
studies of binary compact object formation (e.g., Portegies Zwart \&
Verbunt~1996; Belczynski, Kalogera \& Bulik~2002) indicate that a
small fraction of progenitors might not be circularized before the
explosion.}. We note that the synchronized He star has a large
specific angular momentum compared to a maximal-rotating NS. So the
He-star core has to shed significant angular momentum before it can
collapse to a NS. Such angular momentum loss (most likely mediated by
magnetic stress; see Heger et al.~2004) does not change the spin
orientation of the He core.

We shall make similar reasonable assumption (that the pre-SN binaries
are circular) about MS/He-star binaries (the progenitor systems of
MS/NS binaries) and WD/He-star binaries (the progenitor systems of
MWD/NS binaries).

\begin{table*}[htp!!]
  \begin{center} 
    \caption{Tidal circularization timescale of different types of 
He-star/NS binaries  \label{tab3}}
    \begin{tabular}{lcccccccc} 
      \tableline\tableline
      Case &  $M_{\rm He}$ &  $R_{\rm He}$ & Log($L_{\rm He}/L_{\sun}$) & 
      $\Pb$ & $t_{\rm f}$ & $t_{\rm circ}$ \\
       & $\Msun$ & $R_{\sun}$ &  & days & yr & yr \\
      \tableline
      BB & 2.8 & 0.5 & 4.5 & 0.08 & 0.01 & 170 \\
      BB & 2.8 & 1.8 & 4.5 & 0.5  & 0.03 & 100 \\
      BB & 2.8 & 2.8 & 4.5 & 1.0  & 0.04 & 120 \\
      BC & 2.8 & 12  & 4.5 & 10   & 0.10 & 330 \\
      BC & 3.6 & 2   & 5   & 0.6  & 0.02 & 140 \\
      \tableline
    \end{tabular}
  \end{center}
\end{table*}

Assumption (2): In deriving eqs.~(\ref{eqvk}) and (\ref{eqgamma}), we
have assumed that the SN of star B (He star) does not affect the mass
and motion of star A. As discussed below, in several systems, the
post-explosion spin orientation of star A, $\bS_{Af}$, can be
constrained observationally, and we will assume that the spin
direction of star A is unchanged during the explosion,
i.e. $\bS_{Af}\parallel \bS_{Ai}$.

To justify these assumptions, consider the explosion of the He star in
a MS/He-star binary (the progenitor of a MS/NS system)\footnote{The
effect of ejecta capture by the NS in a NS/He-star binary is obviously
much smaller.}  and we can estimate the ejecta mass and momentum
captured by the MS star. Assuming that the explosion of the He star is
only mildly aspherical (indeed, only small asymmetry is enough to
generate large kick to the NS), the ejecta mass captured by the MS
star is
\begin{eqnarray}
m_{\rm cap} \sim \frac{\pi R_{\rm MS}^2}{4\pi a_{\rm i}^2}m_{\rm ej}
\end{eqnarray}
where $R_{\rm MS}$ is the radius of the MS star, $a_{\rm i}$ the
binary separation before explosion, and $m_{\rm ej}$ is the total mass
ejected by the He star. As an estimate, we use the parameters for the
PSR J0045-7319/MS system (see \S3.3 below): From the observed
semi-major axis $a_0$ and eccentricity $e_0$ of the current PSR/MS
binary, we find $a_{\rm i}>a_{\rm f}(1-e_{\rm f})\simeq
a_0(1-e_0)\simeq 4R_{\rm MS}$. Thus at most $1/64$ of the ejecta mass
is captured by the MS star, or $m_{\rm cap}\la 0.03M_\odot$ for
$m_{\rm ej}=m_{\rm He}-m_{\rm NS}\sim 2M_\odot$.  For a typical mass
ejection velocity $V_{\rm ej}\sim10^4\,\kms$, the MS star will receive
an impact velocity $\Delta V$ less than $30\,\kms$.  In general,
requiring $\Delta V=m_{\rm cap}V_{\rm ej}/m_{\rm MS}$ to be less than
pre-explosion orbital velocity of the MS star, $(Gm_{\rm
tot}/\ai)^{1/2}m_{\rm He}/m_{\rm tot}$ (where $m_{\rm tot}=m_{\rm
MS}+m_{\rm He}$), leads to the condition $\ai\ga 2R_{\rm MS}$.  Thus
the momentum impact on the MS star due to the ejecta is negligible in
most cases. We note that even in extreme cases, when $\ai/R_{\rm
MS}=$a~few, the momentum impact is along the pre-SN orbital
plane. This impact, by itself, does not change the orientation of the
orbital plane.

From the above considerations, we conclude that the assumption
$\bS_{Ai}\parallel \bS_{Bi}\parallel\bL_i\parallel \bS_{Af}$ is most
likely valid. Thus the angle $\theta$ between $\bL_i$ and $\bL_f$ is
equal to the angle between $\bL_f$ and $\bS_{Af}$, which can be
constrained observationally (see below).

What about spin orientation $\bS_{Bf}$ of the newly formed NS (star
B)?  Obviously, if the angular momentum of the NS originates entirely
from its progenitor, then $\bS_{Bf}\parallel\bS_{Bi}$. But it is
possible that even with zero ``primordial'' rotation, finite NS
rotation can be produced by off-centered kicks (Spruit \& Phinney
1998; see also Burrows et al.~1995, who reported a proto-NS rotation
period of order 1~s generated by stochastic torques in their 2D
supernova simulations). Such SN-generated spin will necessarily make
$\bS_{Bf}$ misalign with $\bS_{Bi}$. Thus in general, the angle
$\gamma$ in eq.~(\ref{eqgamma}) refers to the angle between the kick
$\bV_k$ imparted on star B and its primordial spin $\bS_{Bi}$, not
necessarily the angle between $\bV_K$ and $\bS_{Bf}$ (see \S 4).
 
Finally, the post-SN binary may continue to evolve, so the currently
observed orbital elements $a_0,~e_0$ may not be the same as
$\af,~\ef$.  Depending on the type of systems, the evolution may be
driven by gravitational radiation or tidal effect.

%%%%%%%%%%%%%%%%%%%%%%%%%%
\subsection{NS/NS Binaries}

Currently there are 8 observed NS/NS binary systems observed in our
Galaxy (7 are listed in Stairs 2004, plus PSR J1756-2251 reported by
Faulkner et al.~2005)\footnote{Our Table 4 does not include the new
system PSR J1906+0746 ($\Pb=3.98\,h$, $e=0.0853$, $M_{\rm
tot}=2.6\,\Msun$) recently discovered by Arecibo (Lorimer et
al.~2005). With the reasonable assumptions that $\mai=2.1-8.0\,\Msun$,
$\ma=\mb=1.3\,\Msun$ and $\theta=0\arcdeg-180\arcdeg$, we can give the
kick velocity $\Vk=50-1590\,\kms$ and the misalignment between kick
and pre-spin axis $\gamma = 24\arcdeg-156\arcdeg$.} They consist of a
recycled millisecond pulsar (A) and a second-born NS (B) which, in
most cases, is not visible as a radio pulsar (with the exception of
the double pulsar system, PSR J0737-3039).

Following the birth of the second-born NS (B) in a SN, the NS/NS
binary evolution is governed by gravitational radiation reaction, with
(Peters 1964)
\begin{eqnarray}
\frac{1}{a}\frac{{\rm d}a}{{\rm d}t} &=& -\frac{64G^3}{5c^5}\frac{\ma \mb M_f}
{a^4(1-e^2)^{7/2}}\left(1+\frac{73}{24}e^2+\frac{37}{96}e^4\right),
\label{eqda}\\
\frac{1}{e}\frac{{\rm d}e}{{\rm d}t} &=&
-\frac{304G^3}{15c^5}\frac{\ma \mb M_f}
{a^4(1-e^2)^{5/2}}\left(1+\frac{121}{304}e^2\right).\label{eqde}
\end{eqnarray}
The time lapse $T$ since the SN to the present is unknown.  In our
calculations below we adopt the characteristic age of the pulsar,
$\tau_{\rm c}=P/(2\dot P)$, as a conservative upper limit to $T$,
although the actual value of $T$ may be a lot smaller (see below for
specific cases). From the measured $a_0$ and $e_0$, we can integrate
eqs.~(\ref{eqda})-(\ref{eqde}) backward in time to obtain $\af$ and
$\ef$.  We also note that the angle between $\bS_A$ and $\bL_f$ is
unchanged during the evolution and is given by $\theta$.

We now consider individual NS/NS binary systems (see Table 4).  The
kicks in PSR J0737-3039, PSR B1534+12, PSR B1913+16 have been studied
in detail by Willems et al.~(2004) (see also Dewi \& van den Heuvel
2004; Piran \& Shaviv 2004) and PSR B1534+12 by Thorsett et
al.~(2005).  Our constraints on the kick are based on the observed
orbital elements and spin parameters of the systems. For several of
the systems, additional constraint can be obtained from the proper
motion and location of the binary in the Galaxy. We summarize our key
findings below (see Table 4).

\begin{table*}[htp!!!]
  \begin{small}
  \begin{center} 
    \caption{Constraints on kicks in double NS binaries \label{tab4}}
    \begin{tabular}{lcccccccc} 
      \tableline\tableline
      PSR & $\mb$  & $\ma$   & $\mbi$  & $\ef$ & $\af$ & $\theta$ & $\gamma$ & $\Vk$  \\
          &$\Msun$ & $\Msun$ & $\Msun$ &       & ls    & deg      & deg      & $\kms$ \\
      \tableline
J0737-3039      &   1.25 &    1.34 &   2.1--4.7 &   0.088--0.103 &     2.94--3.28   &     0--180  &   24--156                &    60--1610   \\
                &   1.25 &    1.34 &   2.3--3.3 &   0.088--0.103 &     2.94--3.28   &     0--180  &   28--152                &    80--1490   \\
                &   1.25 &    1.34 &   2.3--3.3 &   0.088--0.103 &     2.94--3.28   &    60   &   36--47  or  133--144 &   610--760    \\
\\
J1518+4904      &   1.05 &    1.56 &   2.1--8.0 &            0.249 &              56.68 &     0--180  &   19--161                &    11--500    \\
                &   1.31 &    1.31 &   2.1--8.0 &            0.249 &              56.68 &     0--180  &   12--168                &     4--500    \\
\\
B1534+12        &   1.35 &    1.33 &   2.1--8.0 &   0.274--0.282 &     7.63--7.82   &    21--29   &   10--74  or  106--170 &   160--480    \\
                &   1.35 &    1.33 &   2.1--8.0 &   0.274--0.282 &     7.63--7.82   &   151--159  &   75--83  or   97--105 &   660--1360   \\
\\
J1756-2251      &   1.20 &    1.37 &   2.1--8.0 &   0.181--0.197 &     6.27--6.65   &     0--180  &   20--160                &    31--1390   \\
\\
J1811-1736      &   1.11 &    1.62 &   2.1--8.0 &            0.828 &              96.58 &     0--180  &    0--180                &     0--873    \\
                &   1.37 &    1.37 &   2.1--8.0 &            0.828 &              96.58 &     0--180  &    0--180                &     0--867    \\
\\
J1829+2456      &   1.36 &    1.14 &   2.1--8.0 &   0.139--0.149 &    14.78--15.47  &     0--180  &   20--160                &    19--867    \\
                &   1.25 &    1.25 &   2.1--8.0 &   0.139--0.149 &    14.78--15.47  &     0--180  &   22--158                &    25--870    \\
\\
B1913+16        &   1.39 &    1.44 &   2.1--8.0 &   0.617--0.657 &     6.51--7.44   &    12--24   &    0--84  or   96--180 &   180--660    \\
                &   1.39 &    1.44 &   2.1--8.0 &   0.617--0.657 &     6.51--7.44   &   156--168  &   77--87  or   93--103 &   530--2170   \\
\\
B2127+11C       &   1.36 &    1.35 &   2.1--8.0 &   0.681--0.725 &     6.57--7.81   &     0--180  &    0--180                &     0--2370   \\
\tableline
    \end{tabular}
  \end{center}
  \end{small}
\end{table*}

{\it PSR J0737-3039}. This double-pulsar system contains 22.7~ms
pulsar (Pulsar A) and a 2.77~s pulsar (Pulsar B), with the current
orbital period $\Pb$=0.1\,d and eccentricity $e_0$=0.0878 (Lyne et
al.~2004).  For the He progenitor mass of pulsar B, we consider both
the range $\mbi=2.1-4.7M_\odot$ (for stable mass transfer) as well as
the range $\mbi=2.3-3.3M_\odot$ as suggested by Dewi \& van den Heuvel
(2004). The spindown time of pulsar A is 210 Myr and pulsar B 50~Myr;
we use the latter as the approximate upper limit for the time lapse
since the SN, thus we consider $\ef$=0.0878~--~0.103 and
$\af$=2.94~--~3.28\,$\ls$.  The angle $\theta$ has not be measured
directly.  Theoretical modeling of the eclipse lightcurve (Lyutikov \&
Thompson 2005) suggests that pulsar-B's spin axis $\bS_{Bf}$ is
inclined from the orbital angular momentum vector by 60$\arcdeg$ (or
120$\arcdeg$); if $\bS_{Bf}\parallel \bS_{Bi}$ (see \S 3.1), then this
angle is equal to $\theta$.  Our constraints on $V_K$ and $\gamma$ for
different cases are given in Table 4. For the same input parameters
($\theta=0\arcdeg - 180\arcdeg$, $\mbi=2.1 - 4.7 \Msun$), our result
is consistent with that of Willems et al.~(2004) (see their Table
2). With the more constrained parameters, we obtain stronger
constraints on the kick: for $\theta=60\arcdeg$, the kick direction is
$\gamma=36\arcdeg$--$47\arcdeg$ or $133\arcdeg$--$144\arcdeg$ with
velocity $\Vk=610-760\,\kms$; for $\theta=120\arcdeg$, even larger
kicks ($\Vk=1040-1300\,\kms$) are required\footnote{Such large kick
velocities are likely inconsistent with the small measured proper
motion of the system ($V_T<30$~km/s; see Kramer et al.~2005; Coles et
al.~2005) unless the system has significant line-of-sight
velocity. One may thus rule out the Lyutikov-Thompson solution
($\theta=60\arcdeg$ or $120\arcdeg$), or, more likely, if the the
Lyutikov-Thompson solution is correct, one may conclude that
$\bS_{Bf}$ is not parallel to $\bS_{Bi}$.}.

Additional constraint from the observed proper motion and galactic
location of the system tends to rule out the high values of kick
velocity (see Piran \& Shaviv 2004; Willems et al.~2004), but does not
change the fact that a misaligned kick ($\gamma\neq 0$) is required to
explain the observed orbital property of the system.

{\it PSR J1518+4904} contains a 40.9~ms pulsar with $P_{\rm
orb}=8.634$~d and $e_0=0.249$ (Nice, Sayer \& Taylor 1996; Hobbs et
al.~2004). Thorsett \& Chakrabarty (1999) gave the total mass
$M_f=\ma+\mb=2.62(7)M_\odot$, and $\ma=1.56^{+0.13}_{-0.44}$,
$\mb=1.05^{+0.45}_{-0.11}$. The spindown age of this pulsar is 16~Gyr,
According to eqs.~(\ref{eqda})-(\ref{eqde}), we find that the
eccentricity and semi-major axis have changed very little, $\af\simeq
a_0$, $\ef\simeq e_0$. The spin-orbit inclination angle $\theta$ for
Pulsar A is currently unknown (although it will likely be constrained
by measuring geodetic precession in the next few years), so we
consider the full range $\theta=0-180\arcdeg$.  For
$M_{Bi}=1.05M_\odot$, the kick direction is constrained to
$19\arcdeg-161\arcdeg$ and the kick velocity could be as small as
10\,$\kms$. For more typical masses, $M_{A}=1.32\,\Msun,
M_{B}=1.30\,\Msun$, the kick constraints are similar.

{\it PSR B1534+12} contains a 37.9~ms pulsar with $\Pb=0.42$~d and
$e_0=0.274$.  Measurement of geodetic precession of the pulsar
constrains the spin-orbit inclination angle $\theta=25\arcdeg\pm
4\arcdeg$ or $\theta=155\arcdeg\pm 4\arcdeg$ (Stairs et al.~2004;
Thorsett et al.~2005). The spindown time is 250~Myr, and according to
eqs.~(\ref{eqda})-(\ref{eqde}), we obtain the range
$\ef$=0.274~--~0.282, $\af$=7.63~--~7.82\,$\ls$.  For
$\theta=25\arcdeg\pm4\arcdeg$, our result ($\Vk\simeq160-480\,\kms$)
is consistent with Thorsett et al.~(2005), who used additional
constraint from proper motion ($\mu_{\alpha}=1.34(1)\,\maspyr$,
$\mu_{\delta}=-25.05(2)\,\maspyr$; with the distance of 0.68\,kpc,
this corresponds to transverse velocity of $\VT=107.04\,\kms$; Konacji
et al.~2003) to obtain $\Vk=230\pm60$\,km/s. The
$\theta=155\arcdeg\pm4\arcdeg$ solution leads to much greater values
of $\Vk$ and might be inconsistent with the observed proper motion.
In all cases, aligned kicks ($\gamma=0$) are ruled out.

{\it PSR J1756-2251} contains 28.5~ms pulsar with $\Pb=0.32$~d and
$e_0=0.181$ (Faulkner et al.~2005).  The total system mass
$\ma+\mb=2.574\,M_\odot$ from measurement of relativistic periastron
advance, and the companion mass is constrained to
$\mb<1.25\,M_\odot$. We adopt $\mb=1.2M_\odot$ and $\ma=1.37M_\odot$,
$\theta=0\arcdeg$~--~$180\arcdeg$. The spin down age is 443~Myr.  The
kick direction is constrained in $20\arcdeg-160\arcdeg$, which means
that aligned kicks are ruled out.

{\it PSR J1811-1736} contains a 104.2ms pulsar with $P_{\rm
orb}=18.8$~d and $e_0=0.828$ (see Stairs 2004). The mass ranges are
$\ma=1.62^{+0.22}_{-0.55}$, $\mb=1.11^{+0.53}_{-0.15}$. The spindown
age is about 900~Myr.  For $\theta$=$0\arcdeg$~--~$180\arcdeg$, the
kick is not well constrained.  Aligned kicks are allowed for
relatively low $\mbi$.

{\it PSR J1829+2456} contains a 41~ms pulsar ($\tau_{\rm
c}=12.4\,$Gyr) with $\Pb=1.18$~d and $e_0=0.139$ (Champion et al.~2004;
Champion et al. 2005). The masses are $\ma=1.14^{+0.28}_{-0.48}$ and
$\mb=1.36^{+0.50}_{-0.17}$.  The kicks are constrained with
$\gamma=20\arcdeg-160\arcdeg$ and $V_{\rm k}=20-880\,\kms$, which
means spin-kick misaligned.  For $\ma=\mb=1.25\,\Msun$, the
constraints are similar.

{\it PSR B1913+16} (Hulse-Taylor binary) contains a 59~ms pulsar with
$\Pb=0.323$~d and $e_0=0.617$.  Measurement of geodetic precession of
the pulsar constrains the spin-orbit inclination angle
$\theta=18\arcdeg\pm6\arcdeg$ or $\theta=162\arcdeg\pm6\arcdeg$ (Wex
et al.~2000; Weisberg \& Taylor~2003). The spindown time of the pulsar
is 109 Myr, so we adopt $T<109 Myr$, giving $\ef=0.617-0.657$,
$\af=6.51-7.44$\,ls.  For the pre-SN mass of the He star in the range
of 2.1--8.0$\Msun$, we get a similar result compared to Willems et
al.~(2004) (see their Table 2): if $\theta=18\arcdeg\pm6\arcdeg$,
$\gamma$=$0\arcdeg$~--~$84\arcdeg$ or $96\arcdeg$~--~$180\arcdeg$,
$\Vk\simeq180-660\,\kms$; if $\theta=162\arcdeg\pm6\arcdeg$,
$\gamma$=$77\arcdeg$~--~$87\arcdeg$ or $93\arcdeg$~--~$103\arcdeg$,
$\Vk\simeq530-2170\,\kms$.  The transverse velocity $\VT=88\,\kms$
($\mu_{\alpha}=-2.56(6)\,\maspyr$, $\mu_{\delta}=0.49(7)\,\maspyr$,
see Weisberg \& Taylor~2003; and distance 7.13\,kpc, see Taylor \&
Cordes~1993) suggests that the $\theta=18\arcdeg\pm6\arcdeg$ solution
may be more reasonable.

{\it PSR B2127+11C} contains 30.53~ms-pulsar with $P_{\rm
orb}=0.335$~d, $e_0=0.681$ and characteristic age 97.2\,Myr (Deich \&
Kulkarni~1996). According to eqs.~(\ref{eqda})-(\ref{eqde}), we find
$\ef=0.681-0.725$, $\af=6.57-7.81$\,ls. The kicks are not well
constrained, and aligned kicks are allowed by the data.  Note that
this system lies in the globular cluster M15, and it may have formed
by many-body interactions in the cluster core.

Our general conclusion from NS/NS binaries is that modest kicks which
are misaligned with the pre-SN orbital angular momentum are required
to produce the spin-orbit characteristics of these systems (see also
van den Heuvel 2004).

%%%%%%%%%%%%%%%%%%%%%%%%%%%%%%%%%%%%%%%%%%%%%%%%%%%%%%%%%%%%%%%%%%%%%5
\subsection{Pulsar/MS Binaries}

There are three published pulsar/MS binary systems (see Table 5)
\footnote{We do not include the system PSR J1638-4715, a 0.764~s
pulsar with a Be star companion, with $\Pb\simeq 1800$~d and
$e_0=0.808$ (A. Lyne 2005, private communication).} Such systems
evolve from He-star/MS binaries (which in turn evolves from MS/MS
binaries) when the He star explodes in a SN to form a NS. After the
explosion, the pulsar/MS binary may further evolve under tidal
interaction if the orbit is sufficiently compact. The evidence for
such tidal evolution is most clear in the PSR J0045-7319 system.

\begin{table*}[htp!!!!]
  \begin{center} 
    \caption{Constraints on kicks in pulsar/MS binaries \label{tab5}}
    \begin{tabular}{lcccccccc} 
      \hline\hline
      PSR & $\ma$  & $\mb$   & $\mai$  & $\ef$ & $\af$ & $\theta$ & $\gamma$ & $\Vk$ \\
          &$\Msun$ & $\Msun$ & $\Msun$ &       & ls    & deg      & deg      & $\kms$ \\
      \hline
J0045-7319 &   1.58 &  8.8 &   2.1--8.0 & 0.808 & 293.96 &   115--160  &   53--85  or   95--127 &   119--740  \\
J0045-7319 &   1.58 &  8.8 &   2.1--8.0 & 0.950 & 128.79 &   115--160  &   52--87  or   93--128 &    52--750  \\
J0045-7319 &   1.58 &  11  &   2.1--8.0 & 0.808 & 293.96 &   115--160  &   53--85  or   95--127 &   130--800 \\
\\
B1259-63   &   1.35 &  10  &   2.1--8.0 & 0.870 & 531.51 &    0--180   &    0--180              &    13--330    \\
B1259-63   &   1.35 &  10  &   2.1--8.0 & 0.870 & 531.51 &    10--40   &    0--86  or   94--180 &    22--110    \\
\\
J1740-3052 &   1.35 &  12--20 &   2.1--8.0 & 0.579 & 873.26 &    30--75   &    0--64  or  116--180 &    37--240    \\
J1740-3052 &   1.35 &  12--20 &   2.1--8.0 & 0.579 & 873.26 &   105--150  &   48--80  or  100--132 &    89--370    \\
      \hline
    \end{tabular}
  \end{center}
\end{table*}

{\it PSR J0045-7319} contains a 0.926~s pulsar (with characteristic
age 3 Myr) in an orbit with a B star companion (mass$=10\pm1\,\Msun$)
with $\Pb=51$~d and $e_0$=0.808 (Kaspi et al.~1996).  Timing
observation revealed the effects of classical spin-orbit coupling due
to the rapid rotation of the B star, including periastron advance and
precession of the orbital plane (Kaspi et al.~1996; see also Lai et
al.~1995; Wex 1998). This constrains the angle between $\bL$ and
$\bS_B$ (spin axis of the B star) to be
$\theta=20\arcdeg$~--~$65\arcdeg$ or
$\theta=115\arcdeg$~--~$160\arcdeg$.  Timing data also revealed rapid
orbital decay, $\Pb/\dot \Pb =-0.5$~Myr, which can be naturally
explained by dynamical tidal interaction between the pulsar and the B
star near periastron, provided that the B star rotation axis is
(approximately) opposite to $\bL$ (Lai 1996; Kumar \& Quataert
1997). We therefore choose the $\theta=115\arcdeg$~--~$160\arcdeg$
solution.  Lai (1997) showed that during the dynamical tidal
evolution, $a(1-e)$ is approximately constant. Since tidal interaction
tends to align $\bS_B$ and $\bL$ and the current system has a highly
misaligned geometry, the binary should not have evolved very long. We
therefore consider the range $\ef=0.808-0.95$ (most likely $\ef$ is
rather close to $e_0$). We find that with different values of $\ef$,
the implied $\Vk$ and $\gamma$ change only slightly: $\Vk\simeq
50-750\,\kms$, and $\gamma\simeq52\arcdeg-87\arcdeg$ or
$93\arcdeg-128\arcdeg$, which implies spin-kick misalignment.

Another important point for this system is that the initial spin of
the pulsar can be constrained. A generic theory of dynamical-tide
induced orbital decay combined with the currently measured orbital
decay rate, yield an upper limit, 1.4~Myr, to the age of the system
since the SN (Lai 1996).  Since the characteristic age of the pulsar
is 3~Myr, we find that the initial spin period of the pulsar is close
to the current value, $\Pinit\ga 0.5$~s.

{\it PSR B1259-63} contains a 47.8~ms pulsar and a Be star companion
(mass $\mb\simeq 10\,\Msun$), with $\Pb=1236.7$~d and $e_0$=0.870 (see
Wex et al.~1998; Wang, Johnston \& Manchester 2004).  Because of the
large orbital period and the short characteristic age (0.33\,Myr),
tidal interaction is probably unimportant, i.e., the evolution of the
system since the SN is negligible.  We therefore adopt $\ef\simeq e_0$
and $\af\simeq a_0$.  (With different $\ef$, the constraint on the
kick is similar.)  Although spin-orbital coupling effects are likely
important in this system, no self-consistent timing solution has been
obtained so far (see Wang et al.~2004; Johnston et al.~2005a) and
hence $\theta$ has not been measured. For
$\theta=0\arcdeg$~--~$180\arcdeg$, the kick velocity is constrained to
10~--~340\,$\kms$, and the kick direction is unconstrained. In
particular, for $\theta=23\arcdeg-42\arcdeg$, aligned kicks
($\gamma=0$) are allowed (see Fig.~\ref{fig1}). Melatos et al.~(1995)
modeled the variation of the dispersion measure and rotation measure
near periastron produced by the circumstellar disk of the Be star and
suggested that the disk is tilted with respect to the orbital plane by
$10\arcdeg-40\arcdeg$. For $\theta$ in this range, aligned kicks are
also allowed.

\begin{figure}
\includegraphics[angle=-90,scale=.5]{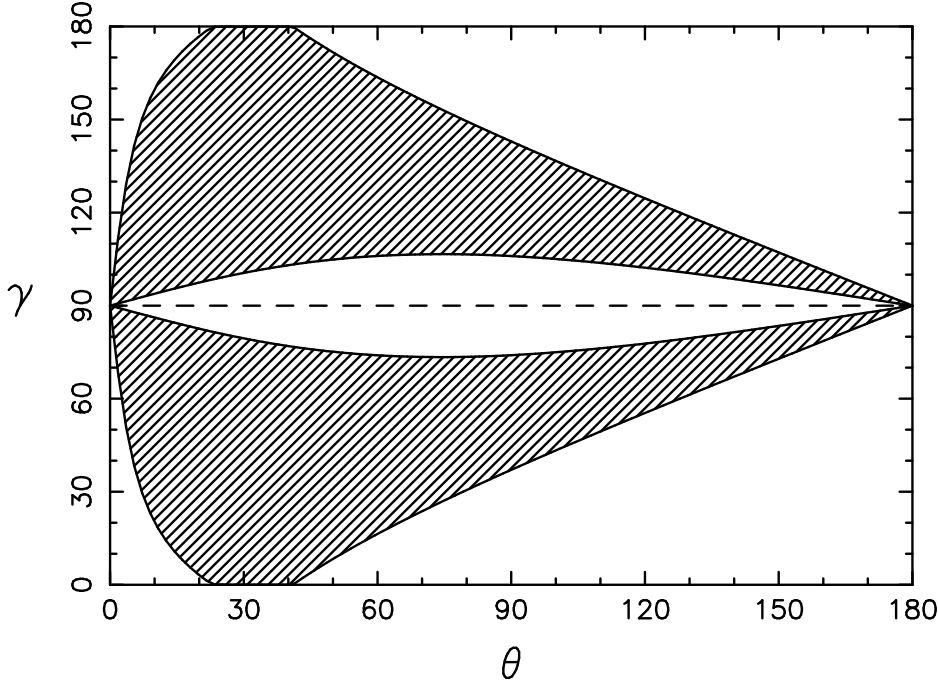}
\caption{The allowed kick angle $\gamma$ (shaded region) 
as a function of spin-orbit misalignment angle $\theta$
in the PSR B1259-63 system.
\label{fig2}}
\end{figure}

{\it PSR J1740-3052} contains a 0.57~s pulsar (with characteristic age
$0.35$~Myr) and an early Be star companion (mass$>11\,\Msun$), with
$\Pb=231$~d and $e_0$=0.579 (Stairs et al.~2001).  The large orbital
period and small characteristic age imply that tidal evolution has not
changed the binary parameters, thus we use $\ef\simeq e_0$.  The
spin-orbital inclination angle $\theta$ lies in the range
$30\arcdeg$~--~$75\arcdeg$ (or $105\arcdeg$~--~$150\arcdeg$) (see
Stairs et al.~2003).  We find that aligned kicks ($\gamma=0$) are
allowed for this system for $\theta=30\arcdeg-75\arcdeg$, but not
allowed for $\theta=105\arcdeg-150\arcdeg$.

%%%%%%%%%%%%%%%%%%%%%%%%%%%%%%%%%%%%%%%%%%%%%%%%%%%%%%%%%%%%%%
\subsection{Young pulsars with massive white dwarf companions}

Such systems are thought to evolve from binaries in which both stars
are initially less massive than the critical mass required to produce
a NS (Tauris \& Sennels 2000; Stairs 2004).  The initially more
massive star transfers mass to its companion before becoming a WD. If
sufficient matter can be accreted by the initially low mass star, it
will exceed the critical mass and produce a NS. Should the system
remain bound, an eccentric binary with a young NS and a MWD companion
will be produced.  Two such systems are known (see Table 6).

\begin{table*}[htp!!!!!]
  \begin{center} 
    \caption{Constraints on kicks in binary pulsars with massive WD companion \label{tab6}}
    \begin{tabular}{lcccccccc} 
      \tableline\tableline
      PSR & $\mb$  & $\ma$   & $\mbi$  & $\ef$ & $\af$ & $\theta$ & $\gamma$ & $\Vk$ \\
          &$\Msun$ & $\Msun$ & $\Msun$ &       & ls    & deg      & deg      & $\kms$ \\
      \tableline
J1141-6545      &   1.30 &    0.99 &   2.1--8.0 &            0.172 &               4.38 &    15--30   &   21--78  or  102--159 &   160--620    \\
                &   1.30 &    0.99 &   2.1--8.0 &            0.172 &               4.38 &   150--165  &   76--85  or   95--104 &   890--1610   \\
\\
B2303+46        &   1.34 &    1.30 &   2.1--8.0 &            0.658 &              72.16 &     0--180  &    0--180                &     0--690    \\
      \tableline
    \end{tabular}
  \end{center}
\end{table*}

{\it PSR J1141-6545} contains a WD (mass=0.986\,$\Msun$) and a young
pulsar ($\Pp$=0.4\,s, mass=1.30\,$\Msun$), with current orbital period
$\Pb$=0.198\,d and $e_0$=0.172 (see Bailes et al.2003). Measurement of
geodetic precession constrains the spin-orbit inclination angle to
$\theta=15\arcdeg-30\arcdeg$ (or $150\arcdeg$~--~$165\arcdeg$) (see
Hotan et al.~2005).  From the pulsar's spindown age (1.46\,Myr), we
find that the orbit has evolved very little, thus $\ef\simeq
e_0=0.172$ and $\af\simeq a_0=4.38$\,ls.  For all reasonable mass of
the NS progenitor, aligned kicks are ruled out (see Table 6).

{\it PSR B2303+46} contains a white dwarf (mass=1.3\,$\Msun$) and a
young pulsar ($\Pp$=1.1\,s, mass=1.34\,$\Msun$), with current orbital
period $\Pb$=12.34\,d and $e_0$=0.658 (see Thorsett et al.~1993; van
Kerkwijk \& Kulkarni 1999).  The spin-orbital inclination angle
$\theta$ has not been measured.  From the characteristic age of the
pulsar (30\,Myr), we find that the orbit has evolved very little after
the NS's birth.  The kick direction for this system is not
constrained.

%%%%%%%%%%%%
\subsection{High-Mass X-Ray Binaries}

A HMXB consists of a NS, which often appears as an X-ray pulsar, and a
massive stellar companion (e.g. a Be star).  Of the $\sim 130$ known
HMXBs, about 20 have well-constrained orbital elements, mostly
determined from the timing of the X-ray pulsars.

\begin{table*}[htp!!!!]
  \begin{center} 
    \caption{Constraints on kicks in some Be X-ray binary systems \label{tab7}}
    \begin{tabular}{lcccccc} 
      \tableline\tableline
      PSR & $\Pb$ & $\ef$ & $\ma$   & $\mb$   & $\mai$  & $\Vk$ \\
          & days  &       & $\Msun$ & $\Msun$ & $\Msun$ & $\kms$ \\ 
      \tableline
4U0115+63    & 24.32   &   0.340 &   1.33 &    8.00 &   2.1--8.0 &     0--470    \\
             & 24.32   &   0.340 &   1.33 &   20.00 &   2.1--8.0 &     3--660    \\
\\
V0332+53     & 34.25   &   0.310 &   1.33 &    8.00 &   2.1--8.0 &     0--410    \\
             & 34.25   &   0.310 &   1.33 &   20.00 &   2.1--8.0 &     0--580    \\
\\
2S1417-624   & 42.12   &   0.446 &   1.33 &    8.00 &   2.1--8.0 &     0--440    \\
             & 42.12   &   0.446 &   1.33 &   20.00 &   2.1--8.0 &    15--620    \\
\\
EXO2030+375  & 46.01   &   0.370 &   1.33 &    8.00 &   2.1--8.0 &     0--360    \\
             & 46.01   &   0.370 &   1.33 &   20.00 &   2.1--8.0 &     5--510    \\
\\
A0535+26     & 110.3   &   0.470 &   1.33 &    8.00 &   2.1--8.0 &     0--320    \\
             & 110.3   &   0.470 &   1.33 &   20.00 &   2.1--8.0 &    13--460    \\
      \tableline
    \end{tabular}
  \end{center}
\end{table*}

There are three classes of HMXBs, distinguished by their orbital
parameters (the first two classes are apparent in Table 3 of Bildsten
et al. 1997): (1) Systems with $\Pb\lesssim10$ days and
$e\lesssim0.1$: The low orbit period and low eccentricity indicate
that tidal circularization has played a significant role. So one
cannot obtain constraint on NS kicks from the observed orbital
parameters.  (2) Be X-ray binaries with moderately wide orbits and
high eccentricities ($\Pb\sim20-100$ days, $e\sim0.3-0.5$). The high
eccentricities indicates these systems have received large kicks, with
a mean speed of $\sim 300\,\kms$ (Table 7; see Pfahl et al.~2002). (3)
Possibly another class of Be X-ray binaries has recently been
identified by Pfahl et al.~(2002) (see their Table 1). These systems
are distinguished from the well-known Be X-ray binaries by their wide
orbits (all have $P_{\rm orb}>30$ days) and fairly low eccentricities
($e\lesssim0.2$).  The NSs born in these systems apparently have
received only a small kick, $\lesssim50\,\kms$. Pfahl et al.~(2002)
and van den Heuvel (2004) discussed possible origin for the bimodal
kicks in the two classes of Be X-ray binaries.

The eccentric Be X-ray binaries have similar orbital parameters as the
pulsar/MS binaries discussed in \S 3.3. In Table 7 we list 5 Be X-ray
binaries which belong to the second class discussed in the prevous
paragraph.  With the reasonable assumptions that the mass of the
companion main-sequence star $\mb=8-20\,\Msun$, the mass of the NS
progenitor $\mai=2.1-8.0\,\Msun$ and the NS mass $\ma=1.33\,\Msun$,
the allowed range of kick velocities is given in Table 7. Since the
angle $\theta$ is unknown, we find the kick directional angle $\gamma$
is unconstrained for these systems, i.e. $\gamma=0\arcdeg-180\arcdeg$
(see Table 7), similar to the PSR B1259-63 case (cf. Table 5). The
magnitude of the kick velocity ranges from 0 to several hundred
$\kms$.

%%%%%%%%%%%%%%%%%%%%%%%%%%%%%%%%%%%%%%%%%%%%%%%%%%%%%%%%%%%%
\section{Discussion}

\subsection{Tentative Inference from Observational data}

Our analysis of the velocity-spin correlation for isolated pulsars (\S
2) shows that kick is aligned with spin axis for many (but most likely
not all) pulsars (see Table 1-2 and Fig.~1).  Of particular interest
is the fact that for pulsars with estimated initial spin periods (when
such estimate can be made) less than $\sim 200$~ms, the kick is
apparently aligned with the spin axis to within the error of
measurements (typically $\pm 10\arcdeg$; see Table 1).  If we exclude
PSR J0538+281 (for which $\Pinit\sim 140$~ms), the initial spin
periods of these pulsars are all less than $\sim 70$~ms. Of course, we
should add the usual caveat that a pulsar might have experienced a
phase of rapid spindown before electromagnetic braking begins, so that
the actual spin period of the proto-NS may be shorter than the
estimated $\Pinit$.

On the other hand, our analysis of SN kicks in NS binaries based on
the observed spin-orbital property of various NS binaries (\S 3) shows
that in a number of systems, the kick must not be aligned with the
spin axis of the {\it NS progenitor}.  How can we reconcile this
conclusion with the apparent spin-kick alignment for many isolated
pulsars?

One possibility is that although the kick $\bV_k$ is misaligned with
the spin axis (denoted by $\bS_{Bi}$ in \S 3) of the He star, it may
still be aligned with the spin axis of the NS, since the NS may get
most of its angular momentum from off-centered kicks rather than from
its He star progenitor (Spruit \& Phinney 1998; see also Burrows et
al.~1995).  However, in the absence of any ``primordial'' angular
momentum from the progenitor, a kick (of any duration) displaced by a
distance $s$ from the center, produces a spin of $\simeq
12\,(\Vk/300~\kms) (s/10~{\rm km})$~Hz, with the spin axis necessarily
perpendicular to the velocity direction.  If the kick is the result of
many thrusts on the proto-NS (Spruit \& Phinney 1998), the relative
direction of the net kick and spin depends on how the orientation of
each thrust is correlated with each other.  Spin-kick alignment is
possible in some circumstances, but is by no means a generic
prediction of the ``multiple thrusts'' scenario.  If the kick
direction is not aligned with the He star spin (as we show for many NS
binaries in \S 3), it will not be aligned with the NS spin in general,
regardless of the origin of the NS spin.

An important clue comes from the PSR 0045-7319/B-star system: The kick
imparted to the pulsar at its birth is misaligned with its spin axis,
and the initial spin period has been constrained to be $\Pinit>0.5$~s
(see \S 3.3). Also, for PSR 0737-3039 (with pulsar B spin period
$P_0=2.77$~s) and many other NS/NS binaries, as well as for the PSR
J1141-6545/MWD system (with $P_0=0.39$~s), the kick is misaligned with
the spin axis. In addition, for the PSR J1740-3053/MS (with
$P_0=0.57$~s) system, the kick and spin may be misaligned.

Combining these kick constraints obtained from NS binaries with the
information we have about kicks in isolated pulsars, we are led to the
following tentative suggestion: When the NS initial spin period is
less than a few $\times 100$~ms, the kick will be aligned the the spin
axis; otherwise, the kick will in general not be aligned with the spin
axis, except by chance.  This suggestion, by no means definitive, is
consistent with all the observational data on NS kicks in isolated
pulsars and in NS binaries which have have analyzed/summarized in \S 2
and \S 3.

We now discuss the implications of this suggestion for 
NS kick mechanisms.

\subsection{Implications for kick mechanisms}

Many mechanisms for NS kicks have been suggested or studied. They
generally fall into the following categories (e.g., 
Lai 2004; Janka et al.~2004).

(i) {\it Hydrodynamically driven kicks} in which the SN
explosion is asymmetric (with the explosion stronger in one direction
than the other directions), and the NS receives a kick according to
momentum conservation. Large-scale convections in the neutrino-heated
mantle behind the stalled shock (at $\sim 100$~km) may naturally lead
to such asymmetric explosion, particularly when the delay between core
bounce and shock revival is sufficiently long to allow for small-scale
convective eddies to merge into bigger ones (Scheck et al.~2004; see
also Thompson 2000; Blondin \& Mezzacappa 2005; Foglizzo et
al.~2005). Pre-SN asymmetric perturbations due to convective O-Si
burning (Bazan \& Arnett 1998), amplified during core collapse (Lai \&
Goldreich 2000), may also play a role (Burrows \& Hayes 1996;
Goldreich et al.~1996; Fryer 2004).  The kick timescale ranges from
10's ms to 100's ms.  Obviously, detailed calculations of this class
of mechanisms are still uncertain --- such a calculation/simulation is
an integral part of the general problem of SN explosion mechanism.

(ii) {\it Magnetic-Neutrino Driven kicks} rely on asymmetric neutrino
emission induced by strong magnetic fields. This could arise because
the strong magnetic field modifies the neutrino opacities either
through standard weak-interaction physics (e.g. Dorofeev et al.~1985;
Lai \& Qian 1998; Arras \& Lai 1999a,b; Duan \& Qian 2005) or through
nonstandard physics (e.g., Fuller et al.~2003; Lambiase~2005). It
could also arise from the dynamical effect of the magnetic field on
the proto-NS (e.g., the B field can affect the neutrino-driven
convection/instabilities, and thus creating dark or hot neutrino
spots; Duncan \& Thompson 1992; Socrates et al.~2004).  All these
effects are important only when the magnetic field of the proto-NS is
stronger than $10^{15}$~G.  The kick timescales are of order the
neutrino diffusion time, a few seconds.

(iii) {\it Electromagnetically driven kicks} involve radiation from
off-centered rotating dipole, which, for appropriate dipole
orientation/displacement, imparts a gradual acceleration to the pulsar
along its spin axis (Harrison \& Tademaru 1975; Lai et al.~2001).
This effect is important only if the NS initial period is $\la
2$~ms. The kick time is of order the initial spin-down time ($\simeq
10^7$~s for $B=10^{13}$~G and $\Pinit=14$~ms).

(iv) Other mechanisms are possible if the collapsing iron core has
large angular momentum. For example, the combination of rapid rotation
and magnetic field may lead to bipolar jets from the SN, and a slight
asymmetry between the two jets will lead to a large kick (e.g.
Khokhlov et al.~1999; Akiyama et al.~2003). Another mechanism could be
that, if a rapidly rotating core fragments into a double proto-NS
binary (current numerical simulation indicates this is unlikely; see
Fryer \& Warren 2003), the explosion of the lighter proto-NS (after
mass transfer) could give the remaining NS a kick (Colpi \& Wasserman
2002; see also Davis et al.~2002).

It is of interest to use observations to constrain or rule out some of
these mechanisms. Since the initial spin period of radio pulsars is
$\gg 1$~ms, (iii) and (iv) appear unlikely in general.  The observed
dipole magnetic field of most radio pulsars lies in the range
$10^{12}-10^{13}$~G, but it is not clear whether most proto-NSs can
have (even transient) magnetic fields stronger than $10^{14}$~G.  So
we cannot easily rule out (ii).

Regarding spin-kick alignment/misalignment, the crucial point is the
ratio between the initial spin period $\Pinit$ and kick timescale
$\tau_{\rm kick}$.  In hydrodynamically driven kicks (i) and
magnetic-neutrino driven kicks (ii), the primary thrust to the NS does
not depend on the NS spin axis. But the net kick will be affected by
rotational averaging if $\Pinit$ is much less than $\tau_{\rm
kick}$. Let $V_0$ be the kick velocity that the NS attains in the case
of zero rotation, and $\theta_k$ be the angle between the primary
asymmetry and the rotation axis. The expected components of kick along
the rotation axis and perpendicular to it are (for $\tau_{\rm kick}\gg
\Pinit$)
\begin{eqnarray}
V_{{\rm kick}\parallel}=V_0\cos \theta_k,\qquad
V_{{\rm kick}\perp}\sim {\sqrt{2}\,P\over 2\pi\,\tau_{\rm kick}}
V_0\sin \theta_k.
\end{eqnarray}
Thus the angle $\gamma$ between the kick vector ${\bf V}_{\rm kick}$
and the spin axis is given by $\tan\gamma\sim 0.2(\Pinit/\tau_{\rm
kick}) \tan\theta_k$. Typically, the spin-kick alignment will be
achieved when $\tau_{\rm kick}\gg \Pinit$.  The observed spin-kick
alignment for $\Pinit\la 100$'s ms discussed in \S 4.1 therefore
suggests that $\tau_{\rm kick}$ lies between hundreds of ms to 1~s.
Such kick timescale is consistent with magnetic-neutrino driven
mechanisms or hydrodynamical mechanisms with long-delayed SN
explosions.

Obviously, this conclusion is far from definitive. For example, since
the primary thrust may be applied at a distance larger than the NS
radius, a somewhat more stringent condition on $\Pinit$ is required to
produce spin-kick alignment (see Lai et al.~2001). In another word,
the inequality $\Pinit\ll \tau_{\rm kick}$ is a necessary (but not
sufficient) condition for spin-kick alignment. Also, an initial spin
period of order $100$'s ms could be generated by the SN kick itself
(see above). We have argued (see \S 4.1) that a kick-induced spin
without ``primordial'' angular momentum (i.e. from the progenitor)
would not in general give rise to spin-velocity alignment.  The
situation may be different with even a modest primordial spin. We
plan to study these issues in the future.

%\clearpage

\acknowledgments

This work is supported by National Natural Science Foundation
of China (10328305, 1025313 and 10473015). DL has also been supported
in part by NSF grant AST 0307252 and NASA grant NAG 5-12034.
DL thanks IAS (Princeton), CITA (Toronto), Tsinghua University
(Beijing) and particularly NAOC (Beijing) for hospitality during the
course of the work.

\end{document}